\documentclass[10pt,twocolumn]{article}
\usepackage[margin=0.75in]{geometry}
\usepackage[T1]{fontenc}
\usepackage{times}
\usepackage{booktabs,array,amssymb,amsmath,xcolor,url,graphicx}
\usepackage[hidelinks]{hyperref}
\usepackage{xurl}
\usepackage{enumitem}
\setlist{nosep,leftmargin=*}
\usepackage{placeins}

\title{\textbf{How Are Multi-Adapter LLM Serving Systems Evaluated?}\\[4pt]
\large An Empirical Study of Baselines, Workloads and Artifacts in 62 Multi-LoRA Serving Papers}
\author{Georgii Kliukovkin\\Independent Researcher\\\texttt{kliukovkin@gmail.com}}
\date{}
\begin{document}
\maketitle

\begin{abstract}
Low-rank adapters turn one base model into thousands of cheap task- or user-specific variants, and serving those variants concurrently on shared accelerators (multi-LoRA serving) has become its own systems problem. This paper is an empirical study of how the systems proposed for it since 2022 are evaluated. We code 62 systems against a fixed 25-field schema in two passes, audit their repositories, and ask what each was compared against, at what scale, on what workload, and whether it can be obtained. Four findings stand out. First, the baseline called ``vLLM'' in 27 papers is at least five different things: the engine's native multi-adapter path (17 papers), adapter weights merged into the model (6), the bare backbone (3), an inference engine beside a separate fine-tuning framework (3), and the authors' own model-swapping baseline (1); three papers run two of these. Throughput gains reported over ``vLLM'' range from a few percent to 57.9$\times$ and, because the baselines differ, cannot be ordered. Second, among the 47 systems that state it, the median main experiment uses 20 adapters; ten report some run with 1{,}000 or more. Third, we found no public request trace of language-model serving that carries adapter identifiers; the one public production dataset we found that records adapter use is for an image-generation service and gives only the number of adapters per request. Fourth, we could inspect a repository or artifact for 24 of 62 systems; the most-used research baseline has not changed since January 2024 and is archived. A census of 7{,}051 public LoRA adapters of seven base models gives a median rank of 16 and shows 1\% of adapters receiving 30\% of downloads. We release the coded corpus and the census and close with what a shared benchmark for this area must fix.
\end{abstract}

\section{Introduction}

Low-Rank Adaptation (LoRA)~\cite{ax210609685} freezes a base model and trains a pair of small matrices per layer, so a fine-tuned variant costs megabytes instead of gigabytes. The consequence for inference is structural: an operator no longer serves a handful of models but one base model with hundreds to millions of variants attached to it. Per-user personalization~\cite{ax240204401}, per-task adapter collections~\cite{ax240500732}, adapters generated on demand from a task description or a document~\cite{ax250606105,ax260215902}, and reinforcement-learning loops that continuously produce new adapter revisions~\cite{ax260513779,ax260508527} all push the number of variants up while the traffic per variant goes down.

Serving one model well does not solve this. A merged adapter runs at base-model cost but pins a full copy of the weights; an unmerged adapter shares the base but adds work to every token and breaks the assumption that all requests in a batch execute the same weights. Adapters must be located, loaded, kept resident or evicted; requests for different adapters and ranks compete inside one batch; KV caches computed under one adapter are, in general, invalid under another. Punica~\cite{ax231018547} and S-LoRA~\cite{ax231103285} framed the problem in late 2023. By November 2026 our corpus holds 62 systems, 31 of them with a 2026 publication or preprint date.

Existing surveys do not cover this area in depth. General surveys of LLM serving~\cite{ax231215234,ax240712391,ax250419720,ax250621901,ax250501658,ax261033475,ax260417227} and surveys of LoRA and parameter-efficient fine-tuning~\cite{ax240711046,ax250100365,ax240314608,ax260421905} give adapter serving between nothing and roughly a thousand words and name a handful of systems each (Section~\ref{sec:related}). Meanwhile each new system compares itself with one or two predecessors on its own workload, so the field has no shared picture of what has been built, what composes with what, or which results can be compared.

\paragraph{Research questions and contributions.} The unit of analysis is the evaluation section and the artifact of each paper, and the results are counts over a coded corpus of 62 systems, each described by the same 25 fields, 52 of them checked in a second pass. We ask: (RQ1) what is each system compared against, and in which configuration; (RQ2) at what scale and on what workload is it evaluated; (RQ3) can it be obtained; (RQ4) what has been built, and for which adapters does it work.
\begin{itemize}
\item A \textbf{baseline analysis}: what each baseline called ``vLLM'' or ``S-LoRA'' actually was, which papers state a version, and why headline gains cannot be ordered (Section~\ref{sec:eval}, Table~\ref{tab:eval}, Figure~\ref{fig:vllm}).
\item A \textbf{scale and workload analysis}: adapter counts, trained versus synthetic adapters, the sources of request workloads, and what is publicly known about adapter popularity (Section~\ref{sec:eval}).
\item A \textbf{reproducibility audit}: which systems publish code, in what state, and what they are built on (Section~\ref{sec:repro}).
\item A \textbf{census of public adapters}: ranks, target modules, file sizes and download skew of 7{,}051 adapters of seven base models (Section~\ref{sec:eval}).
\item A \textbf{lifecycle index} of the corpus used to organize the counts, with compatibility and exactness tags (Section~\ref{sec:tax}, Table~\ref{tab:systems}).
\end{itemize}
We measure no system ourselves. The question here is what the literature claims and whether those claims can be compared or checked; measuring current engines against each other is separate work.

\section{Background}\label{sec:bg}

\paragraph{Adapter arithmetic.} For a frozen weight $W\in\mathbb{R}^{d\times k}$, LoRA learns $B\in\mathbb{R}^{d\times r}$ and $A\in\mathbb{R}^{r\times k}$ with $r\ll\min(d,k)$ and computes $h = Wx + BAx$. \emph{Merged} serving folds $BA$ into $W$ and costs nothing extra per token, but one copy of $W$ then serves one adapter. \emph{Unmerged} serving keeps $W$ shared and evaluates $BAx$ on the side, which costs $O(r(d+k))$ extra operations per adapted matrix per token and lets requests for different adapters share a batch. Almost every design decision in this area is a way to move along the line between these two points, or to make the unmerged side cheaper.

\paragraph{Why the base-model serving stack is not enough.} Modern engines batch continuously, manage KV cache in pages~\cite{ax230906180} and split work between prefill and decode~\cite{ax240302310,ax231118677}. Adapters interact with each of these. (i) A batch now contains rows that need different $(A,B)$ pairs of different ranks, so the adapter path needs gather-style kernels. (ii) Adapter weights and KV cache compete for the same accelerator memory, and adapter load time appears in time-to-first-token. (iii) A prefix cached under adapter $i$ cannot be reused under adapter $j$, which defeats prefix caching in multi-agent pipelines. (iv) Popularity over adapters is skewed and shifts, so placement and routing have to be adapter-aware.

\paragraph{Other variant formats.} Prefix and prompt tuning~\cite{ax210100190} add virtual tokens and therefore pay in KV cache instead of weights; (IA)$^3$~\cite{ax220505638}, DoRA~\cite{ax240209353} and VeRA~\cite{ax231011454} change the per-token arithmetic; full fine-tunes can be stored as compressed deltas against the base~\cite{ax231205215,ax240210193}. We treat all of these as \emph{variants of a shared base} and include systems that serve them.

\section{Methodology}\label{sec:method}

\paragraph{Search and selection.} The corpus was assembled in two steps. We first collected candidates by keyword search over arXiv, USENIX, ACM, IEEE, MLSys, NeurIPS, ICML, ICLR and IJCAI proceedings and OpenAlex, and by following references forward and backward from Punica, S-LoRA and dLoRA. We then ran nine fixed queries against the arXiv API (\emph{multi-LoRA} with \emph{serving} or \emph{inference}; \emph{LoRA adapters} with \emph{serving} and \emph{latency} or \emph{throughput}; \emph{multi-tenant LoRA serving}; \emph{adapter serving}; \emph{LoRA serving}; \emph{LoRA} with \emph{cold start}; \emph{delta}, \emph{fine-tuned}, \emph{serving}, \emph{compression}; \emph{KV cache} with \emph{LoRA} and \emph{agents} or \emph{sharing}; \emph{co-serving}, \emph{fine-tuning}, \emph{inference}), restricted to submissions since January 2022. The queries returned 153 hits, 120 unique. Thirty-seven had already been coded (34 in scope, three training-only), seven more were judged in scope, and 76 were excluded on title and abstract. Of the seven, four entered the corpus; the other three are a delta format without a multi-variant experiment~\cite{ax251219720}, an evaluation of an existing server~\cite{ax240500732} and a workload characterization~\cite{ax260923321}, which we cite but do not count as systems. The fixed queries recover 34 of the in-scope papers that reference chasing had already found (55 of the 62 systems in the final table come from reference chasing), so they are a check on the corpus and not a census. We included a paper if it addresses a systems problem of executing, placing, loading, caching or managing \emph{many variants} of a shared base model at inference time and reports a systems metric. We excluded training-only methods and purely algorithmic quality work.

\paragraph{Coding.} Each paper was coded against a fixed schema of 25 fields (problem, mechanism, lifecycle stage, compatibility, exactness and reported quality cost, base models, number and rank of adapters, whether adapters are trained or synthetic, workload generation, trace availability, baselines with their configuration and version, hardware, metrics, headline claim, implementation substrate, code link). A second pass re-read the latest version of 52 of the 62 papers and checked every field (of the other ten, seven were coded in a single pass and three were added later, as described below). It changed at least one field in about nine records in ten, most often the headline claim, the adapter count and the baseline list, which is why we report only fields that survived the second pass; the corrections are part of the data release. Three records had been coded from a superseded version of the paper and were re-coded. Two records (Rock, Dop) were read through a tool that summarizes pages. Three records (MixLoRA, DyMerge-LoRA, CLIMB) were added on 24 November 2026, after the second pass, and were coded once by the author from the publisher or OpenReview pages; DyMerge-LoRA was coded from four of its eleven pages (pages 1, 4, 6 and 7: setup, baselines and main results) and CLIMB from its OpenReview forum page, pages 5--6 of the PDF and its repository README. Extraction in both passes was assisted by a large language model; the final table was curated and checked by the author. \emph{Main} adapter count means the count in the paper's primary end-to-end comparison; where a paper sweeps the count we take the default it names, or the largest point of the main figure if it names none. We code a system as \emph{exact} if its outputs are unchanged relative to serving the same adapters unmerged, and as \emph{drop-in} if it serves existing adapters or existing full-model deltas without retraining.

\paragraph{Corpus.} The result is \textbf{62 systems} (Table~\ref{tab:systems}): 38 peer-reviewed (36 at conferences and 2 in journals), 2 at workshops and 22 available only as preprints, some of them weeks old. Throughout, ``peer-reviewed'' means a conference or journal paper, and the two strata we compare are peer-reviewed papers and preprints together with workshop papers. The year we give is that of the version we coded. Pairs of preprint and published versions are counted once and coded from the published version: CaraServe/Toppings, FastLibra/ELORA, and LoRA-Switch/AdaFuse, whose AAAI version arXiv marks as having substantial text overlap with the LoRA-Switch preprint and which shares its authors and mechanism~\cite{ax240517741,ax260311873}. Seven papers that address only multi-adapter \emph{training}~\cite{ax231202515,ax250802932,ax251000206,ax260207263,ax260405426,ax250901193,ax260302885} are outside the corpus. Eleven systems are boundary cases that concern a single model with internal adapter experts or routing among its own adapters, a few adapters on a device, hardware, or co-serving of one adapter with training (AdaFuse~\cite{ax260311873,ax240517741}, MoLoRA, PRIMAL, CacheReforge, QUAD, OneForAll-LLM, SHiRA, FlexLLM, LayerLoom, MARLaaS, DeltaServe); we keep them because they change what a multi-variant server must do. Three coded papers were excluded as not serving systems (CA-LoRA, a foundation-model technical report, and a single-adapter confidential-computing study). Seven further papers were identified but not coded: six were behind a paywall or bot protection when we tried (Cannikin~\cite{cannikin}, IEEE TPDS; AuLoRA, ICCD; UniLoRA and LoRATEE, ICASSP; LoRA-PIM, ISOCC; CoLoRA, ASP-DAC), and one non-peer-reviewed preprint (FlashServe) we read only in summarized form. All six are IEEE or ACM publications, so the corpus under-represents those venues.

\paragraph{Sensitivity.} We recomputed the findings under alternative choices. Counting the two artifacts that are linked but that we did not open as available raises 24 of 62 to 26. Counting Symbiosis's single vLLM run and MinT's own compatibility test as vLLM baselines would raise the vLLM count by two, from 27 papers to 29. The median main adapter count is 20 over all 47 systems that state one, 32 over the 38 that remain without the eleven boundary cases, 20 among the 25 peer-reviewed papers (conference or journal) and 15 among the 22 preprints and workshop papers; we report the overall figure.

\paragraph{Limits.} We report what papers state. We did not run any system; how far the five baseline configurations differ on one workload is a measurement this study does not provide. We did not measure the accuracy of the coded fields after the second pass. Ten records had no second pass: the seven coded in a single pass and the three added on 24 November. Version and adapter-provenance fields are lower bounds, because fifteen papers were not read to the end of their appendices. Repository facts come from the hosting site on 23 November 2026, and for five systems from pages opened on 24--25 November 2026; we did not build or run any repository. All statements about a paper refer to the text we read.

\section{Reproducibility audit}\label{sec:repro}

\paragraph{Code.} For every system we looked for a repository: we followed links in the paper and, where there was none, searched by system name and title. For 24 of the 62 systems we could inspect a repository or artifact record of the system's own. For 19 of them we recorded the full set of fields; for five (BitDelta, Delta-CoMe, FlexLLM, CacheReforge and CLIMB) we opened the repository page, README or archive record on 24--25 November 2026 but did not obtain a last-push date, and none of the first four is archived. Two more papers link code that we did not open: PetS links an archived artifact on Zenodo that we did not open (a third-party mirror of it exists), and DyMerge-LoRA links an archive record that carries the ACM ``Artifacts Available'' badge. Two have only a branch on a fork or a draft pull request against an engine. For the remaining 34 we found no code; at least five of those papers state that code will be released. Of the 19 fully audited, two are archived: S-LoRA, the most-used research baseline, and Punica, last pushed in January and May 2024. Ten had a push in the twelve months before our check. Of the 14 whose README we could classify, four document how to start an HTTP inference server; the others provide kernels, scripts or experiment harnesses. For dLoRA there is an artifact snapshot from May 2024; for Toppings the proceedings show artifact badges but we found no repository.

\paragraph{Substrate.} Twenty-five of the 62 prototypes are built on vLLM, three on SGLang, five on S-LoRA or its LightLLM base, six on Hugging Face Transformers\slash PEFT, one on llama.cpp and 18 on custom runtimes or simulators; three do not state it and one uses another runtime. The share built on vLLM or SGLang was 2 of 10 among systems dated 2024, 9 of 20 for 2025 and 17 of 31 for 2026. This is good for adoption and bad for comparison, because ``built on vLLM'' spans versions from 0.1.3 to 0.30 in our corpus, and over that period the engine's own multi-adapter path went from absent to one whose kernels its source attributes to Punica and which keeps adapters in an LRU cache.

\paragraph{Consequence.} A new system that wants to compare against a published one usually cannot run it and compares instead against the engine it forked. The citation graph of baselines (Section~\ref{sec:eval}) reflects exactly this.

\section{How systems are evaluated}\label{sec:eval}

Table~\ref{tab:eval} lists, for every system that has either a serving-system baseline or a request workload, what it was compared with and on what.

\begin{table*}[tp]
\centering\scriptsize
\caption{Evaluation practice. \emph{Built on}: implementation substrate. \emph{Baselines}: serving systems compared against; only serving systems count as baselines, so fine-tuning frameworks, serverless platforms, ablations and quality-only baselines are not listed. The superscript on vLLM gives its configuration in that paper: n~native multi-adapter path, m~adapter weights merged into the model (a copy per adapter, or merge and unmerge on demand), b~backbone only, s~authors' model-swapping baseline on vLLM, p~inference engine beside a separate fine-tuning framework on split resources. \emph{Ver.}: a version or commit of a baseline is stated. \emph{Main}/\emph{Max}: adapters in the main end-to-end evaluation and the largest number in any experiment. \emph{Adapters} in serving experiments: trained or downloaded (real), random weights (syn), both (mixed), not stated (n/s). \emph{Workload}: syn.~synthetic arrivals and popularity; AzFn~Azure Functions traces with functions mapped to adapters; LMSYS~Chatbot Arena logs with models mapped to adapters; LLM~tr.~lengths or arrivals from an LLM service trace without adapter identifiers; prod.$^\dagger$~production trace with real adapter usage, not released; prod.$^\ddagger$~production trace re-labelled into more adapters, not released; prod.$^\S$~production request trace without described adapter popularity.}\label{tab:eval}
\setlength{\tabcolsep}{5pt}
\begin{tabular}{@{}l l p{5.6cm} c r r l l@{}}
\toprule
System & Built on & Baselines & Ver. & Main & Max & Adapters & Workload\\
\midrule
PetS & custom & -- &  & 128 & 4344 & n/s & syn. \\
dLoRA & vLLM & vLLM$^{m}$, HF PEFT, S-LoRA &  & 128 & 128 & n/s & AzFn \\
FLoRA & vLLM & -- &  & n/s & n/s & n/s & syn. \\
LoRA-Inlaid & n/s & vLLM$^{m}$, S-LoRA &  & n/s & 1000 & n/s & syn. \\
Punica & custom & vLLM$^{b}$, HF PEFT & \checkmark & n/s & n/s & syn & syn. \\
S-LoRA & S-LoRA & vLLM$^{m}$, HF PEFT &  & 200 & 2000 & n/s & syn., LMSYS \\
SwiftDiffusion & HF & -- &  & 3 & 3 & real & prod.$^\dagger$ \\
Agullo-DT & vLLM & -- &  & n/s & 384 & real & syn. \\
aLoRA & vLLM & vLLM$^{n}$ &  & 5 & 5 & syn & syn. \\
aLoRA-KV & vLLM & vLLM$^{n}$ & \checkmark & 1 & 5 & syn & syn. \\
BD-LoRA & vLLM & S-LoRA, vLLM$^{n}$ & \checkmark & 1 & 128 & real & -- \\
Chameleon & S-LoRA & S-LoRA &  & 100 & 500 & n/s & LLM tr. \\
Compress-then-Serve & vLLM & vLLM$^{n}$ &  & 1024 & 1024 & syn & syn. \\
DeltaZip & vLLM & vLLM$^{s,n}$ &  & 32 & 32 & n/s & syn., AzFn \\
Dop & custom & Punica, vLLM$^{b}$, HF PEFT &  & n/s & n/s & n/s & syn. \\
EdgeLoRA & llama.cpp & llama.cpp &  & 50 & 1000 & n/s & syn. \\
ExpertWeave & vLLM & vLLM$^{m,b}$ & \checkmark & 20 & 20 & real & syn. \\
LLMStation & vLLM & vLLM$^{p}$ &  & n/s & 8 & n/s & LLM tr. \\
Loquetier & HF & HF PEFT, FlexLLM, S-LoRA &  & 4 & 4 & n/s & LLM tr., syn. \\
LoRAServe & S-LoRA & S-LoRA, Toppings &  & 100 & 200 & n/s & prod.$^\ddagger$, LLM tr. \\
MixLoRA & custom & Punica &  & n/s & n/s & n/s & syn. \\
Predictive-LoRA & vLLM & vLLM$^{m}$, S-LoRA, dLoRA &  & 1000 & 1000 & syn & AzFn \\
Rock & HF & -- &  & n/s & 200 & mixed & prod.$^\dagger$ \\
ServerlessLoRA & custom & vLLM$^{m,n}$, dLoRA &  & 10 & 170 & real & AzFn \\
Symbiosis & HF & -- &  & 8 & 32 & n/s & LLM tr., syn. \\
Toppings & S-LoRA & S-LoRA, dLoRA &  & 200 & 512 & syn & AzFn, syn. \\
V-LoRA & other & S-LoRA, Punica, dLoRA &  & n/s & n/s & real & LLM tr. \\
Agullo-Place & vLLM & dLoRA, LoRAServe & \checkmark & 32 & 1280 & real & LLM tr., syn. \\
CLIMB & vLLM & -- &  & 8 & 8 & syn & syn. \\
DeltaServe & vLLM & LLMStation, vLLM$^{p}$ & \checkmark & 1 & n/s & real & prod.$^\S$ \\
DyMerge-LoRA & n/s & vLLM$^{n}$, HF PEFT, S-LoRA &  & n/s & n/s & syn & syn. \\
ELORA & vLLM & vLLM$^{n}$, S-LoRA &  & 100 & 2000 & real & AzFn, LMSYS \\
EpiLoRA & vLLM & vLLM$^{n}$, SGLang, TensorRT-LLM & \checkmark & 32 & 32 & syn & LLM tr., syn. \\
FlexLLM & custom & vLLM$^{p}$ &  & 1 & 1 & n/s & LLM tr. \\
ForkKV & SGLang & vLLM$^{n}$, SGLang & \checkmark & 8 & 8 & n/s & syn. \\
HALO & SGLang & S-LoRA &  & 4 & 4 & n/s & LLM tr. \\
ICaRus & vLLM & vLLM$^{n}$ &  & 8 & 8 & real & syn. \\
InfiniLoRA & vLLM & S-LoRA &  & 512 & 512 & syn & LLM tr., syn. \\
LayerLoom & custom & FlexLLM &  & 1 & 1 & n/s & LLM tr. \\
LoRAForge & SGLang & SGLang, vLLM$^{n}$ & \checkmark & 85 & 85 & real & syn. \\
LoRazor & vLLM & vLLM$^{n}$, S-LoRA & \checkmark & 128 & 128 & real & syn. \\
LRAgent & custom & -- &  & 3 & 3 & real & syn. \\
M-LoRA & S-LoRA & vLLM$^{n}$, Punica, S-LoRA &  & 20 & 200 & n/s & LMSYS \\
MinT & vLLM & -- &  & 64 & n/s & n/s & syn. \\
MoLoRA & custom & -- &  & 4 & 64 & n/s & syn. \\
PLoRA-NDP & custom & S-LoRA &  & 1000 & 1000 & n/s & syn., LMSYS \\
PReCache & vLLM & -- &  & 3 & 8 & real & syn. \\
PreFT & vLLM & vLLM$^{n}$ &  & 32 & 512 & syn & syn. \\
SALT & vLLM & vLLM$^{n}$ &  & 512 & 512 & n/s & -- \\
Scute & vLLM & vLLM$^{n}$ & \checkmark & 1024 & 1024 & syn & syn. \\
\bottomrule
\end{tabular}
\end{table*}

\paragraph{Baselines.} Thirty-nine systems compare against at least one serving system; 23 do not. vLLM is a baseline in 27 of the 39 and S-LoRA in 16; HF PEFT follows with 6, dLoRA with 5, Punica with 4 and SGLang with 3. No other system is a baseline in more than two papers. The comparison graph is a star around two nodes, one of which (vLLM) is a moving target and the other (S-LoRA) an archived repository last pushed in January 2024. Eleven of the 39 state a version or commit for a baseline: four of the 25 peer-reviewed papers and seven of the 14 preprints and workshop papers. Only serving systems count as baselines; schedulers, fine-tuning frameworks and serverless platforms used as further comparison points are not listed.

\begin{figure}[t]
\centering
\includegraphics[width=\columnwidth]{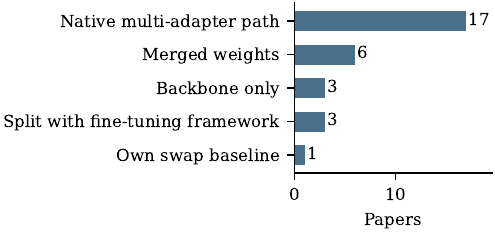}
\caption{Papers per configuration of the baseline called ``vLLM'' (27 papers). Papers that run two configurations are counted in both, so the bars sum to 30.}\label{fig:vllm}
\end{figure}

\paragraph{What ``vLLM'' means.} The second pass recorded how each baseline was configured. Among the 27 papers that compare against vLLM, the name covers at least five configurations (Figure~\ref{fig:vllm}); we count every configuration a paper runs, so the counts sum to 30. In 17 papers it is the engine's native multi-adapter path; in one of them (DyMerge-LoRA) each combination of adapters is first pre-merged into one higher-rank adapter that is then served through that path. In 6 the adapter weights are merged into the model, either as one merged copy per adapter in separate instances or processes that are packed or swapped, or by merging and unmerging on demand (S-LoRA, dLoRA, LoRA-Inlaid, ExpertWeave, Predictive-LoRA, ServerlessLoRA). In 3 it is the backbone alone (Punica; Dop, whose text also says adapters are attached through PEFT and is ambiguous; and a second baseline in ExpertWeave, which uses the vLLM-Ascend plugin). In 3 co-serving papers it is the inference engine beside a separate fine-tuning framework on split resources (FlexLLM, LLMStation, DeltaServe). In DeltaZip it is the authors' own model-swapping baseline built on vLLM. ServerlessLoRA and DeltaZip also run the native path. None of this is a criticism of the individual papers: the early systems predate any multi-adapter path in the engine and chose the only baselines available, and co-serving and full-model-delta papers answer different questions. The problem is the shared name, which lets a reader compare numbers that were never comparable. The same holds for ``S-LoRA'': it is the released system in some papers, a re-implementation or a variant integrated in vLLM in others (InfiniLoRA; Toppings for its 70B runs), a compatibility port in LoRazor, and its tensor-parallel sharding scheme alone in BD-LoRA.

\paragraph{Headline claims.} Reported gains over ``vLLM'' cannot be put on one scale. Punica reports up to 12$\times$ throughput over HF PEFT, DeepSpeed, FasterTransformer and vLLM; switching cost was omitted for all baseline systems, and FasterTransformer and vLLM were run backbone-only, relaxations in the baselines' favour. Against merged copies, S-LoRA reports up to 4$\times$ with few adapters (the baseline runs out of memory beyond that) and dLoRA up to 57.9$\times$ throughput on a 70B model. Against the native multi-adapter path, LoRazor reports up to 3.02$\times$ request throughput on Mistral-7B with 128 adapters built from four trained ones (vLLM 0.30; 1.04--1.12$\times$ at four adapters), Compress-then-Serve about 1.6$\times$ at 1{,}024 adapters (its Figure~1), and EpiLoRA 1.05--1.09$\times$ over the engine's tuned configuration (vLLM 0.28, 32 tenants, on seven of nine GPU--model pairs). In other metrics the range is wider still: the vLLM implementation of aLoRA~\cite{ax251217910} reports up to 58$\times$ lower latency for the adapter-evaluation step. Against S-LoRA, InfiniLoRA reports on average a 3.05$\times$ higher serviceable request rate, LoRAServe up to 2$\times$ throughput, Predictive-LoRA 1.52$\times$ at 1{,}000 adapters, and Chameleon 80.7\% lower P99 TTFT at its highest load point (14.7\% and 24.6\% at lower loads); Toppings reports up to 1.7$\times$ lower latency than S-LoRA and dLoRA. The spread says nothing about which system is better. It means that the systems cannot be ordered from their papers.

\paragraph{Scale.} We recorded two counts: the number of adapters in the main end-to-end evaluation and the largest number in any serving experiment. Forty-seven systems state the first: the median is 20, 29 use 32 or fewer, and four use 1{,}000 or more. The median is 32 without the boundary cases of Section~\ref{sec:method}, 20 among peer-reviewed papers and 15 among preprints and workshop papers. (For DeltaZip the count is of full-model deltas; for DyMerge-LoRA and MixLoRA we record none, because the papers give a slot count and two concurrent ranks, not a number of distinct adapters.) Fifty-one state the second: the median is 50 and ten reach 1{,}000 or more. (We do not count MinT's catalog of a million adapter revisions, of which 64 are warmed in its serving tests.) The headline scale of a paper is thus usually a separate scalability run. Large counts are often obtained with random-weight adapters or by replicating a few trained ones; ELORA and Agullo et al., which use downloaded adapters at their largest scale, are exceptions.

\paragraph{Adapters.} In serving experiments, 23 systems use trained or downloaded adapters, 12 synthetic weights only, one both, and 26 do not say. For throughput the distinction is often immaterial; for any approximate technique, and for anything that depends on adapter content such as compression or cache reuse, it is not.

\paragraph{Workloads.} We found no public request trace of language-model serving that carries adapter identifiers. The abstract of Cannikin~\cite{cannikin}, whose full text we could not read, reports a characterization of a real-world multi-LoRA serving trace; we do not know whether that trace is released. The categories below overlap. Thirty-three systems generate arrivals and adapter popularity synthetically in at least one experiment, 24 of them in all experiments (Poisson or Gamma arrivals with Zipf or power-law popularity, with exponents chosen by the authors). Six map functions of the Azure Functions traces to adapters and four map the models of Chatbot Arena logs to adapters. Twelve take lengths or arrivals from LLM service traces that have no adapter dimension~\cite{ax240117644,ax231118677} and assign adapters synthetically. Four papers use production data. SwiftDiffusion and Rock evaluate on production data from image-generation services with real adapter usage; SwiftDiffusion says a release is being prepared. LoRAServe uses a production trace that contains five adapters and re-labels its requests into 50--200 adapters by a power law; it states that code and traces will be released on acceptance. DeltaServe uses a production request trace without describing adapter popularity. Fourteen systems have no request workload.

\paragraph{What is known about real adapter popularity.} The one public production dataset with adapter information that we found is GenTD26~\cite{gentd26}, from a commercial image-generation service; its maintainers list Rock among its users. Its request file gives, for each request, a timestamp, an anonymized base model and the number of adapters used, but not which adapters, and a characterization study~\cite{ax260923321} reports 874 adapters over 24 days with the top 5\% of adapters accounting for 53.5\% of occurrences, from adapter records that cannot be joined to request timestamps. For language models the evidence is thinner. LoRAServe reports that, of more than a thousand adapters deployed for one model in one region, the top five receive more than 70\% of requests. InfiniLoRA takes its Zipf exponent from production observations in Cannikin, which we could not read. A production LLM trace released in 2026~\cite{maastrace} holds no request-level records: according to its schema it gives daily and five-minute records per serving instance and hourly metrics per application, and marks each instance segment as serving a base model, a fine-tuned model or both. It therefore shows how common fine-tuned variants are in one platform, not how requests distribute over them. We read the documentation and schema of both datasets and the first rows of the GenTD26 request file on 25 November 2026; we did not download the full data.

\paragraph{A census of public adapters.} To put numbers on what a realistic adapter collection looks like, we listed on 24 November 2026 the repositories on the Hugging Face Hub that are tagged as adapters of seven widely used base models (Llama-3.1-8B-Instruct, Llama-3.2-3B-Instruct, Llama-3.3-70B-Instruct, Mistral-7B-Instruct-v0.3, Qwen3-8B, Qwen3-32B and the mixture-of-experts model Qwen3-30B-A3B) and read the adapter configuration file of each. Of 8{,}530 repositories, 7{,}077 have a readable configuration at the repository root, 7{,}051 of them a LoRA configuration with a valid rank; 1{,}246 have no configuration at the root and 207 are gated. The median rank is 16. Ranks 16, 32, 64 and 8 account for 40.8\%, 20.8\%, 19.0\% and 12.0\% of adapters, and 2.1\% have a rank above 128. Ranks rise with the size of the base model: the median is 64 for the 32B model. Three quarters of the adapters (75.4\%) target all seven linear projections of a transformer block, and 9.5\% target only the query and value projections. Adapter files are usually stored in 32-bit floats: the median file for an 8B base model is 160~MiB, about twice its size in half precision in accelerator memory, which matters for loading from disk or network. Downloads over the preceding 30 days are skewed, with 1\% of adapters receiving 30\% of downloads and 10\% receiving 51\%; a rank-frequency fit over the 500 most downloaded adapters has slope $-0.63$ for Llama-3.1-8B-Instruct and $-0.84$ for Qwen3-8B. ServerlessLoRA reports, for a snapshot of 156{,}438 adapter repositories, that the top 10\% account for 94.2\% of downloads~\cite{ax250514468}; our figure is for adapters of seven base models with a readable configuration and for downloads in the preceding 30 days, so the two are not directly comparable. Synthetic workloads in the corpus draw adapter popularity from Zipf or power-law distributions with exponents chosen by the authors (1.2 in InfiniLoRA, taken from a production observation). Repository downloads are not request traffic, and we do not claim that these exponents are wrong; downloads are one more public indicator, alongside the production figures quoted above. The census has further limits: the adapter tag is set by repository authors, ten authors contribute 32\% of the sample, checkpoints of one training run and duplicates are not removed, and it is one snapshot of seven base models. An earlier study of adapter merging collected nearly 1{,}000 public adapters of one of these base models~\cite{recycling}. The census is first reported here; a fuller analysis is planned as part of separate work. The listing and the scripts are part of the data release.

\section{What was built: the corpus by lifecycle stage}\label{sec:tax}

We index systems by the stage of a request's life at which they intervene (Table~\ref{tab:systems}); most touch two or three stages. Orthogonal properties are recorded as tags.

\begin{table*}[tp]
\centering\scriptsize
\caption{The 62 coded systems. Stages: F~format of the variant, S~adapter selection, R~residency and loading, X~execution, N~node-level scheduling, C~cluster placement and routing, K~cross-request (KV-cache) state, L~adapter lifecycle and co-serving with training. Compat.: D~drop-in for existing adapters, T~adapters must be (re)trained in a special form, A~base model or hardware architecture changes, --~does not apply. Ex.: E~exact, $\approx$~approximate (a quality cost is reported), --~not applicable. Base: LLM~dense decoder, MoE, VLM, Diff~diffusion, Enc~encoder. Code: \checkmark~own repository or artifact snapshot, inspected; ($\circ$)~code linked by the paper but not inspected, or only a fork branch or draft pull request; $^{a}$~archived. Yr is the year of the version coded; venue ``--'' is a preprint, W a workshop.}\label{tab:systems}
\setlength{\tabcolsep}{1.6pt}
\begin{minipage}[t]{0.495\textwidth}\centering
\begin{tabular}[t]{@{}l c l *{8}{c} c c l c@{}}
\toprule
System & Yr & Venue & F&S&R&X&N&C&K&L & Cp & Ex & Base & Code\\
\midrule
PetS~\cite{pets} & 22 & ATC & $\bullet$ &  &  & $\bullet$ & $\bullet$ &  &  &  & D & E & Enc & ($\circ$) \\
BitDelta~\cite{ax240210193} & 24 & NeurIPS & $\bullet$ &  & $\bullet$ & $\bullet$ &  &  &  &  & D & $\approx$ & LLM & \checkmark \\
Delta-CoMe~\cite{ax240608903} & 24 & NeurIPS & $\bullet$ &  & $\bullet$ & $\bullet$ &  &  &  &  & D & $\approx$ & LLM & \checkmark \\
dLoRA~\cite{dlora} & 24 & OSDI &  &  &  & $\bullet$ & $\bullet$ & $\bullet$ &  &  & D & E & LLM & \checkmark \\
FLoRA~\cite{ax231205677} & 24 & ICLR &  &  &  & $\bullet$ &  &  &  &  & T & $\approx$ & LLM &  \\
LoRA-Inlaid~\cite{lorainlaid} & 24 & NeurIPS & $\bullet$ &  &  & $\bullet$ & $\bullet$ &  &  &  & D & $\approx$ & LLM & \checkmark \\
ME-Switch~\cite{ax240609041} & 24 & -- & $\bullet$ & $\bullet$ & $\bullet$ & $\bullet$ &  &  &  &  & D & $\approx$ & LLM &  \\
Punica~\cite{ax231018547} & 24 & MLSys &  &  & $\bullet$ & $\bullet$ & $\bullet$ &  &  &  & D & E & LLM & \checkmark$^{a}$ \\
S-LoRA~\cite{ax231103285} & 24 & MLSys &  &  & $\bullet$ & $\bullet$ & $\bullet$ &  &  &  & D & E & LLM & \checkmark$^{a}$ \\
SHiRA~\cite{ax240613175} & 24 & NeurIPS & $\bullet$ &  & $\bullet$ &  &  &  &  &  & T & $\approx$ & Diff & \checkmark \\
SwiftDiffusion~\cite{ax240702031} & 24 & -- &  &  & $\bullet$ & $\bullet$ & $\bullet$ &  &  &  & D & $\approx$ & Diff &  \\
Agullo-DT~\cite{ax250808343} & 25 & NeurIPS-W &  &  &  &  & $\bullet$ &  &  &  & D & E & LLM & \checkmark \\
aLoRA~\cite{ax250412397} & 25 & NeurIPS & $\bullet$ &  &  &  &  &  & $\bullet$ &  & T & E & LLM & \checkmark \\
aLoRA-KV~\cite{ax251217910} & 25 & TMLR &  &  &  & $\bullet$ &  &  & $\bullet$ &  & T & E & LLM & ($\circ$) \\
BD-LoRA~\cite{ax251023346} & 25 & NeurIPS & $\bullet$ &  &  & $\bullet$ &  &  &  &  & T & $\approx$ & LLM & ($\circ$) \\
Chameleon~\cite{ax241117741} & 25 & MICRO &  &  & $\bullet$ &  & $\bullet$ &  &  &  & D & E & LLM &  \\
Compress-then-Serve~\cite{ax240700066} & 25 & ICML & $\bullet$ &  & $\bullet$ &  &  &  &  &  & D & $\approx$ & LLM &  \\
DeltaZip~\cite{ax231205215} & 25 & EuroSys & $\bullet$ &  & $\bullet$ & $\bullet$ &  &  &  &  & D & $\approx$ & LLM & \checkmark \\
Dop~\cite{dop} & 25 & AAAI &  &  &  & $\bullet$ &  &  &  &  & D & E & LLM &  \\
EdgeLoRA~\cite{ax250701438} & 25 & MobiSys &  & $\bullet$ & $\bullet$ & $\bullet$ &  &  &  &  & D & $\approx$ & LLM & \checkmark \\
ExpertWeave~\cite{ax250817624} & 25 & -- & $\bullet$ &  & $\bullet$ & $\bullet$ &  &  &  &  & A & E & MoE &  \\
LLMStation~\cite{llmstation} & 25 & ATC &  &  &  &  & $\bullet$ &  &  & $\bullet$ & -- & E & LLM & \checkmark \\
Loquetier~\cite{ax251100101} & 25 & NeurIPS &  &  &  & $\bullet$ &  &  &  & $\bullet$ & D & E & LLM & \checkmark \\
LoRAServe~\cite{ax251122880} & 25 & -- &  &  & $\bullet$ &  & $\bullet$ & $\bullet$ &  &  & D & E & LLM &  \\
MixLoRA~\cite{mixlora} & 25 & ICPP &  &  &  & $\bullet$ & $\bullet$ &  &  &  & D & E & LLM &  \\
Predictive-LoRA~\cite{ax251220210} & 25 & ICCSMT &  &  & $\bullet$ &  & $\bullet$ &  &  &  & D & E & LLM &  \\
Rock~\cite{rock} & 25 & CLUSTER &  &  & $\bullet$ &  & $\bullet$ & $\bullet$ &  &  & D & E & Diff &  \\
ServerlessLoRA~\cite{ax250514468} & 25 & -- &  &  & $\bullet$ &  & $\bullet$ & $\bullet$ &  &  & D & E & LLM &  \\
Symbiosis~\cite{ax250703220} & 25 & SoCC &  &  &  & $\bullet$ &  &  &  & $\bullet$ & D & E & LLM &  \\
Toppings~\cite{ax240111240} & 25 & ATC &  &  & $\bullet$ & $\bullet$ &  & $\bullet$ &  &  & D & E & LLM &  \\
V-LoRA~\cite{ax241100915} & 25 & EuroSys & $\bullet$ &  &  & $\bullet$ & $\bullet$ &  &  &  & T & $\approx$ & VLM & \checkmark \\
\bottomrule
\end{tabular}
\end{minipage}\hfill
\begin{minipage}[t]{0.495\textwidth}\centering
\begin{tabular}[t]{@{}l c l *{8}{c} c c l c@{}}
\toprule
System & Yr & Venue & F&S&R&X&N&C&K&L & Cp & Ex & Base & Code\\
\midrule
AdaFuse~\cite{ax260311873,ax240517741} & 26 & AAAI & $\bullet$ &  &  & $\bullet$ &  &  &  &  & A & E & LLM &  \\
Agullo-Place~\cite{ax260224044} & 26 & FGCS &  &  &  &  & $\bullet$ & $\bullet$ &  &  & D & E & LLM & \checkmark \\
CacheReforge~\cite{ax260930884} & 26 & -- &  &  &  &  &  &  & $\bullet$ & $\bullet$ & D & $\approx$ & LLM & \checkmark \\
CLIMB~\cite{climb} & 26 & ICML &  &  & $\bullet$ &  & $\bullet$ &  &  &  & D & E & LLM & \checkmark \\
DeltaServe~\cite{ax260728848} & 26 & -- &  &  &  &  & $\bullet$ &  &  & $\bullet$ & D & E & LLM &  \\
DyMerge-LoRA~\cite{dymergelora} & 26 & KDD &  &  & $\bullet$ & $\bullet$ & $\bullet$ &  &  &  & D & E & LLM & ($\circ$) \\
ELORA~\cite{ax250503756} & 26 & HPCA &  &  & $\bullet$ &  &  &  & $\bullet$ &  & D & E & LLM &  \\
EpiLoRA~\cite{ax261103899} & 26 & -- &  &  &  & $\bullet$ &  &  &  & $\bullet$ & D & E & LLM &  \\
FlexLLM~\cite{ax240218789} & 26 & NSDI &  &  &  &  &  &  &  & $\bullet$ & -- & -- & LLM & \checkmark \\
ForkKV~\cite{ax260406370} & 26 & -- &  &  &  &  &  &  & $\bullet$ &  & D & $\approx$ & LLM &  \\
HALO~\cite{ax261004956} & 26 & -- &  &  &  & $\bullet$ & $\bullet$ &  &  &  & D & E & LLM &  \\
ICaRus~\cite{ax260313281} & 26 & ICLR &  &  &  & $\bullet$ &  &  & $\bullet$ &  & T & $\approx$ & LLM &  \\
InfiniLoRA~\cite{ax260407173} & 26 & ATC &  &  & $\bullet$ & $\bullet$ &  & $\bullet$ &  &  & D & E & MoE &  \\
LayerLoom~\cite{ax261100652} & 26 & -- &  &  &  &  &  &  &  & $\bullet$ & -- & -- & LLM &  \\
LoRA-on-a-Budget~\cite{ax261036386} & 26 & -- & $\bullet$ &  &  &  &  &  &  &  & D & $\approx$ & LLM &  \\
LoRAForge~\cite{ax261101256} & 26 & -- & $\bullet$ &  &  & $\bullet$ &  &  &  &  & D & $\approx$ & LLM & \checkmark \\
LoRazor~\cite{ax261110049} & 26 & -- &  &  &  & $\bullet$ &  &  &  &  & D & $\approx$ & LLM &  \\
LRAgent~\cite{ax260201053} & 26 & ICML &  &  &  &  &  &  & $\bullet$ &  & T & $\approx$ & LLM & \checkmark \\
M-LoRA~\cite{mlora26} & 26 & IJCAI &  &  &  &  & $\bullet$ &  &  &  & D & E & LLM &  \\
MARLaaS~\cite{ax260508527} & 26 & -- &  &  &  &  &  &  &  & $\bullet$ & -- & -- & LLM &  \\
MinT~\cite{ax260513779} & 26 & -- & $\bullet$ &  & $\bullet$ &  &  &  &  & $\bullet$ & D & E & MoE & \checkmark \\
MoLoRA~\cite{ax260315965} & 26 & ICML &  & $\bullet$ & $\bullet$ & $\bullet$ &  &  &  &  & D & E & LLM &  \\
OneForAll-LLM~\cite{ax260418655} & 26 & ACL & $\bullet$ &  &  & $\bullet$ &  &  &  &  & A & $\approx$ & LLM &  \\
PLoRA-NDP~\cite{ax260805483} & 26 & -- &  &  & $\bullet$ & $\bullet$ &  &  &  &  & A & E & LLM &  \\
PReCache~\cite{ax260934054} & 26 & -- &  &  &  &  &  &  & $\bullet$ &  & D & $\approx$ & LLM &  \\
PreFT~\cite{ax260514217} & 26 & -- & $\bullet$ &  &  & $\bullet$ &  &  &  &  & T & $\approx$ & LLM & \checkmark \\
PRIMAL~\cite{ax260113628} & 26 & ISCAS &  &  &  & $\bullet$ &  &  &  &  & A & -- & LLM &  \\
QUAD~\cite{ax260329535} & 26 & CVPR-W & $\bullet$ &  & $\bullet$ &  &  &  &  &  & T & $\approx$ & Diff &  \\
RouteKV~\cite{ax260917109} & 26 & -- &  &  &  &  &  &  & $\bullet$ &  & D & $\approx$ & LLM & \checkmark \\
SALT~\cite{ax260803579} & 26 & -- & $\bullet$ &  & $\bullet$ &  &  &  &  &  & T & $\approx$ & LLM &  \\
Scute~\cite{ax261101089} & 26 & -- & $\bullet$ &  &  & $\bullet$ &  &  &  &  & D & E & LLM & \checkmark \\
\bottomrule
\end{tabular}
\end{minipage}
\end{table*}

\subsection{Stage 1: Format of the variant}
Twenty-one systems act here. Some compress a collection of existing adapters and are drop-in but approximate, such as Compress-then-Serve~\cite{ax240700066}, which jointly diagonalizes many LoRAs into a shared basis. Others define formats designed for serving that require adapters trained in that form, such as SALT~\cite{ax260803579}, with a pinned shared centroid and per-user residuals, and BD-LoRA~\cite{ax251023346}, whose block-diagonal factors need no extra tensor-parallel communication. Beyond LoRA, Scute~\cite{ax261101089} maps DoRA, (IA)$^3$, VeRA and OFT onto one normal form and serves them exactly in a batch (its abstract also reports a mean error of at most 1.5\% against PEFT), and DeltaZip~\cite{ax231205215} serves compressed full-model deltas beside a shared base.

\subsection{Stage 2: Adapter selection}
Almost every system assumes that the request names its adapter. Three make the choice part of serving: EdgeLoRA~\cite{ax250701438}, ME-Switch~\cite{ax240609041} and MoLoRA~\cite{ax260315965}, which routes per token so that one pass serves all adapters. A large body of work on retrieving, routing among and composing adapters~\cite{ax240209997,ax240511157,ax230713269} is evaluated for quality, not for what selection costs a serving system. AdaFuse~\cite{ax260311873,ax240517741} merges the selected adapters into the backbone for each token with a fused kernel; it concerns one model with internal experts and not many independent tenants.

\subsection{Stage 3: Residency and loading}
Twenty-six systems decide where adapters live and how fast they appear. S-LoRA~\cite{ax231103285} keeps all adapters in host memory and pages the active ones into a pool shared with the KV cache; later systems add caches, prefetching and pre-loading. CLIMB~\cite{climb} sits in front of vLLM and, when the active adapters outnumber the engine's resident slots, queues requests for non-resident adapters outside the engine, with priority for VIP traffic.

\subsection{Stage 4: Execution}
Execution is the most populated stage (35 systems). Punica's SGMV~\cite{ax231018547} and S-LoRA's rank-heterogeneous kernels gather per-row adapters inside one batch. MixLoRA~\cite{mixlora} runs one worker per LoRA rank on separate CUDA streams so that mixed-rank requests execute concurrently; DyMerge-LoRA~\cite{dymergelora} serves requests that each compose several adapters by fusing them on the GPU at run time. Other systems switch between merged and unmerged modes (dLoRA~\cite{dlora}), stop applying an adapter partway through a request (LoRazor~\cite{ax261110049}, PreFT~\cite{ax260514217}, both approximate), or split the work across servers, devices or proposed hardware (InfiniLoRA~\cite{ax260407173}, PRIMAL~\cite{ax260113628}).

\subsection{Stages 5--6: Scheduling, placement and routing}
On a node (21 systems), schedulers order requests by adapter rank and predicted length~\cite{ax241117741,ax240111240,lorainlaid}, by memory demand with speculation~\cite{mlora26}, or by SLO slack across several services~\cite{ax261004956}. Across nodes, only seven systems act: dLoRA~\cite{dlora}, LoRAServe~\cite{ax251122880}, Toppings~\cite{ax240111240}, InfiniLoRA~\cite{ax260407173}, ServerlessLoRA~\cite{ax250514468}, Rock~\cite{rock} and Agullo et al.~\cite{ax260224044}. Fairness between tenants, studied for base-model serving~\cite{ax240100588}, is touched among the coded systems only by HALO, which schedules several adapter services by SLO slack and compares against fairness schedulers. CLIMB gives one class of traffic priority over another, which is a different goal.

\subsection{Stage 7: Cross-request state}
Nine systems, all dated 2025--2026, address KV-cache reuse across adapters, driven by multi-agent pipelines in which several adapters read the same long context. Some split the cache into a shared base part and a small per-adapter part (ForkKV~\cite{ax260406370}, PReCache~\cite{ax260934054}); others change the adapter or the model so that the base cache is valid by construction (aLoRA~\cite{ax250412397,ax251217910}, ICaRus~\cite{ax260313281}). RouteKV~\cite{ax260917109}, a name taken from its repository (the paper, arXiv:2609.17109, does not use it), studies two adapters on a 1.7B-parameter model and measures what happens if an unmodified adapter takes over a base-model prefix. ELORA~\cite{ax250503756} manages adapters and KV blocks jointly without changing either. Apart from it, every technique at this stage is either approximate or requires retraining.

\subsection{Stage 8: Adapter lifecycle and co-serving}
Ten systems place inference next to the process that produces adapters: six co-schedule fine-tuning and inference on the same base (for example FlexLLM~\cite{ax240218789} and DeltaServe~\cite{ax260728848}), MinT~\cite{ax260513779} and MARLaaS~\cite{ax260508527} build managed platforms around reinforcement-learning loops, and CacheReforge~\cite{ax260930884} and EpiLoRA~\cite{ax261103899} handle continually updated adapters or rollouts. Multi-adapter \emph{training} systems~\cite{ax231202515,ax251000206,ax260302885} are adjacent and outside our counts.

\subsection{Tags and what they show}
\paragraph{Compatibility.} Of 62 systems, 42 are drop-in, 11 require adapters trained in a special form, 5 change the base architecture or hardware, and for 4 co-serving systems the question does not apply. The distinction matters in practice: an operator with an existing collection of adapters can adopt only the first group.

\paragraph{Exactness.} Thirty-three systems are exact and 25 approximate (4 not applicable). The approximate ones report quality on different tasks and metrics, so their costs cannot be ranked against each other.

\paragraph{Base model.} Fifty-three systems evaluate dense decoder LLMs; three target mixture-of-experts models, four diffusion models, one a vision-language model and one encoders.

\paragraph{Environment and trust.} Edge and on-device serving appears in four systems~\cite{ax250701438,ax260418655,ax260329535,ax240613175}. Among the coded systems trust appears as a design goal once (activation privacy in split execution~\cite{ax250703220}); the literature on backdoored adapters in sharing ecosystems~\cite{ax240300108,ax231200374} and on per-permission adapters~\cite{ax250511557} has not been connected to the serving path.

\section{Research and production engines}\label{sec:prod}

Table~\ref{tab:engines} summarizes what open-source engines and orchestration layers document for multi-adapter serving, read from their public documentation and, for vLLM, source on 23 November 2026. We list the engines whose documentation describes adapter handling; the KServe row reflects its LLMInferenceService documentation only (documentation version 0.21; latest release v0.20.0 of 6 August 2026).

\begin{table}[tbp]
\centering\scriptsize
\caption{Multi-adapter support documented by engines and orchestration layers (November 2026).}\label{tab:engines}
\begin{tabular}{@{}p{1.75cm}p{6.3cm}@{}}
\toprule
Engine / layer & Documented multi-adapter support\\
\midrule
vLLM~\cite{docvllm} & Runtime load and unload endpoints; limits on resident GPU and CPU adapters and on rank; requests for different adapters run in parallel up to the resident limit; source attributes its LoRA kernels to Punica and keeps adapters in an LRU cache\\
SGLang~\cite{docsglang} & Runtime load and unload; batching ``based on S-LoRA and Punica'' with two kernel backends; cap on adapters per batch and on loaded adapters; LRU or FIFO eviction; pinning\\
NIM / TensorRT-LLM~\cite{docnim} & NIM: dynamic adapter discovery with a refresh interval, limits as in vLLM. TensorRT-LLM: adapter named per request, cached after first use\\
llama.cpp & Adapter and scale selectable per request; requests with different adapter configurations are not batched together\\
TGI~\cite{doctgi} & Adapters declared at start-up; repository archived, maintenance mode\\
LoRAX~\cite{doclorax} & Just-in-time adapter loading; heterogeneous continuous batching with SGMV\\
Ray Serve LLM~\cite{docray} & Adapters loaded on demand; per-replica LRU with a configurable limit\\
KServe~\cite{dockserve} & Adapters declared on the inference service and loaded at start-up; no dynamic loading\\
AIBrix~\cite{docaibrix,ax250403648} & Adapter as a cluster resource; dynamic loading through a runtime sidecar; choice of adapter scheduler\\
Gateway API Inf.\ Ext. & LoRA-affinity filter in the endpoint picker; adapter roll-out through configuration\\
NVIDIA Dynamo & Runtime endpoints to load, list and remove adapters; adapter resource definition; limits on adapters and rank\\
SageMaker~\cite{docsm} & Adapter inference components attached to a base component\\
\bottomrule
\end{tabular}
\end{table}

Three observations follow from setting this table beside Table~\ref{tab:systems}. First, what reached production is the first generation: gather kernels, paging of adapters with an LRU-style cache, and replica-level routing. Second, formats from Stages~1 and~7 that require retraining have not reached these engines, with the partial exception of aLoRA, which is supported in a training library and was implemented in a vLLM fork~\cite{ax251217910}. Third, the cluster stage in production is handled by orchestration layers whose policies (route to the replica that already holds the adapter) are comparable to the locality routing of ServerlessLoRA and Rock and simpler than the other five research systems at that stage, and none of the seven reports a comparison against them in the text we read. 

\section{Gaps and open problems}\label{sec:open}

\paragraph{Empty cells.} Reading Table~\ref{tab:systems} by columns shows what has not been built. (i) \emph{Selection as a serving problem}: three systems, although selection, retrieval and composition are where much adapter research goes. (ii) \emph{Cluster-level systems}: seven, none reporting a comparison against the routers of production orchestration layers. (iii) \emph{Heterogeneous formats}: Scute (exact, with a mean error of at most 1.5\% against PEFT reported in its abstract) and PetS batch several adapter types and Symbiosis mixes PEFT methods across clients; no system combines format heterogeneity with cache sharing or disaggregation. (iv) \emph{Mixture-of-experts, diffusion and vision-language bases}: three, four and one system respectively, eight in total. (v) \emph{Isolation and admission}: among the coded systems one schedules adapter services against SLOs, one (CLIMB) admits requests for non-resident adapters, and none addresses performance isolation between tenants or admission of untrusted adapters. Three of the papers we could not code address parts of this according to their abstracts (CoLoRA fairness, UniLoRA isolation, LoRATEE trusted execution).

\paragraph{Composability.} Techniques at different stages are presented as independent but are not. Formats that need retraining (aLoRA, PreFT, BD-LoRA, SALT) cannot be applied to an existing adapter collection; approximate execution (LoRazor) and approximate cache reuse (ForkKV, PReCache) each report a small quality cost that has never been measured in combination; compression changes ranks and therefore the behaviour of rank-aware schedulers. Beyond compression paired with a kernel designed for it, we found no evaluation of two such techniques together.

\paragraph{What a shared benchmark must fix.} The meta-analysis points to five requirements: a public language-model workload with adapter identifiers, or at least a published popularity and churn model calibrated on one; trained adapters at realistic rank mixes (the census in Section~\ref{sec:eval} gives one) instead of random weights; named and pinned baseline configurations, including the engine's own multi-adapter path, instead of the bare word ``vLLM''; scale points that separate the regimes (tens, hundreds, thousands of adapters; resident versus cold); and a common quality check for approximate techniques. Methodological guidance from base-model serving on metrics and load generation applies directly~\cite{ax240707000,ax260524217,ax240505465}. Run-to-run reproducibility of the engine itself is a further concern: a November 2026 preprint reports numerical nondeterminism in the kernel that vLLM uses for LoRA multiplications and proposes a deterministic replacement~\cite{detsplitk}.

\paragraph{Adapter churn.} Platforms that mint adapters continuously~\cite{ax260513779,ax260508527} make adapter \emph{revisions}, not adapters, the unit of serving. Hand-off latency, staleness of cached state~\cite{ax260930884} and rollback are barely studied.

\section{Related work}\label{sec:related}

\paragraph{Evaluation practice in LLM serving.} Agrawal et al.~\cite{agrawal25} examine recent LLM inference systems and identify recurring evaluation anti-patterns in baseline fairness, evaluation setup and metric design. Anwar and Keshvadi~\cite{anwar26} note that benchmarks of serving engines rarely control for configuration differences and show, for vLLM and SGLang, how much of a reported gap one matched setting removes. Pape et al.~\cite{pape26} find that the inference stack is rarely reported in machine-learning publications and that the choice of backend alone shifts benchmark scores. Chandrasekar and Kramberger~\cite{ax260524217} identify and mitigate systemic measurement bias in production LLM inference benchmarks. These studies concern base-model serving. We ask the same question of one sub-area, per paper, with a coded table: what each multi-adapter system was compared with, at what scale and on what workload.

\paragraph{Surveys.} Surveys of LLM serving cover batching, scheduling, memory and cluster management for base models; Miao et al.~\cite{ax231215234} and Li et al.~\cite{ax240712391} do not discuss adapter serving, Pan and Li~\cite{ax250621901}, Zhen et al.~\cite{ax250419720}, Noor et al.~\cite{ax261033475} and Xu et al.~\cite{ax260417227} give it between a sentence and a paragraph, and Park et al.~\cite{ax250501658} note adapter support engine by engine. Surveys of LoRA and PEFT are organized around training: Han et al.~\cite{ax240314608} give about a thousand words to PEFT system design with three serving systems, Li et al.~\cite{ax260421905} about five hundred, and Mao et al.~\cite{ax240711046} and Yang et al.~\cite{ax250100365} a paragraph each. A concurrent review by Luk~\cite{luk} studies decentralised peer-to-peer adapter inference; our subject is centralized serving. Surveys of KV-cache management~\cite{ax241219442} treat adapters only as a cache-validity key.

\section{Conclusion}

Multi-adapter serving has grown from a handful of systems first posted in 2022 and 2023 to more than sixty, and in our corpus their results are mostly not comparable with one another. Arranged by request lifecycle, the techniques cover every stage, and one property separates them in practice: whether a technique serves the adapters that already exist. But we could inspect code for 24 of 62 systems and found none for 34, the most-used research baseline has not changed since January 2024, we found no public adapter-level workload for language models, so each paper constructs its own, and the most common baseline name covers five different configurations. We argue that the area would gain more from a public adapter-level trace and a benchmark that holds baselines still than from another kernel.

\paragraph{Data availability.} The coded corpus (one record per system with the second-pass corrections), the repository audit, the census listing with its scripts, and the scripts that produce the tables are available at \url{https://github.com/kliukovkin/multi-lora-serving-evaluation-study}.

\paragraph{Use of AI tools.} The author used a large language model assistant (Claude, Anthropic) for literature search, for extracting the coded fields from the surveyed papers, and for drafting parts of the text. The author checked the final table and takes full responsibility for the content.

\paragraph{Competing interests.} The author is employed by Meta Platforms. This work was done independently, in a personal capacity and on personal equipment; it does not use employer resources or data and does not represent the views of the employer. The author is an open-source contributor to KServe, one of the orchestration layers listed in Table~\ref{tab:engines}, including its LoRA adapter support; that row is taken from public documentation only. The author received no funding for this work.

{\footnotesize
\bibliographystyle{plain}
{\sloppy\bibliography{refs}}
}
\end{document}